\documentclass[epj]{svjour}
\usepackage{graphics}
 
\begin{document} 
\newcommand{\lvecp}{\raisebox{1.77ex}{\mbox{\tiny$\leftarrow$}}\hspace{-0.7em} \partial} 
\newcommand{\p}{\partial} 
\newcommand{\ls}{\left(} 
\newcommand{\rs}{\right)} 
\newcommand{\beq}{\begin{equation}} 
\newcommand{\eeq}{\end{equation}} 
\newcommand{\beqa}{\begin{eqnarray}} 
\newcommand{\eeqa}{\end{eqnarray}} 
\newcommand{\Gs}{\Gamma_s} 
\newcommand{\Go}{\Gamma_0} 
\newcommand{\dGo}{\frac{ \partial{\Gamma_{0}} }{ \partial{\rho_0} }} 
\newcommand{\dGs}{\frac{ \partial{\Gamma_{s}} }{ \partial{\rho_0} }} 
\newcommand{\ddGo}{\frac{ \partial^2 {\Gamma_{0}} }{ \partial{\rho_0^2} }} 
\newcommand{\ddGs}{\frac{ \partial^2 {\Gamma_{s}} }{ \partial{\rho_0^2} }} 
\newcommand{\jj}{(j^{\lambda} j_{\lambda})} 
\newcommand{\dnj}{(\partial^{\nu}j^{\alpha})j_{\alpha}} 
\newcommand{\dr}{(\partial^{\mu} \rho_s)} 
\newcommand{\dnr}{(\partial^{\nu} \rho_s)} 
\newcommand{\su}{\sum_{j=1}^{A \cdot N}} 
\newcommand{\ex}{e^{ \frac{R_j^2}{\sigma^2}}} 
\newcommand{\uj}{(u_j^{\alpha} j_{\alpha})}

\title{Modelization of the EOS} 

\author{C. Fuchs\inst{1} \and H.H. Wolter\inst{2}}
\institute{Institut f\"ur Theoretische Physik der Universit\"at T\"ubingen,  
D-72076 T\"ubingen, Germany \and Sektion Physik der Universit\"at M\"unchen, 
D-85748 Garching, Germany } 
\date{Received: date / Revised version: date}
%
\abstract{This article summarizes theoretical  predictions for the density and 
isospin dependence of the nuclear mean field and the 
corresponding nuclear equation of state. We compare predictions 
from microscopic and phenomenological approaches. An application 
to heavy ion reactions requires to incorporate these 
forces into the framework of dynamical transport models. 
Constraints on the nuclear equation 
of state derived from finite nuclei and from 
heavy ion reactions are discussed.  
} 

\maketitle

\section{Introduction} 
Heavy ion reactions provide the only possibility to reach nuclear 
 matter densities beyond saturation density $\rho_0 \simeq 0.16~{\rm 
 fm}^{-3}$. Transport calculations indicate that in  
the low and intermediate energy range  
$E_{\rm lab}\sim 0.1\div 1$ AGeV nuclear densities between $2\div 3 \rho_0$  
are accessible while the highest baryon densities ($\sim 8 \rho_0$)   
will probably be reached in the energy range of  
the future GSI facility FAIR between $20\div 30$  
AGeV. At even higher  
incident energies transparency sets in and the matter becomes less baryon  
rich due to the dominance of meson production. The isospin dependence of 
the nuclear forces which is at present only little 
constrained by data will be explored by the forthcoming radioactive beam 
facilities at FAIR/GSI \cite{sis200}, SPIRAL2/GANIL and RIA \cite{ria}. 
Since the knowledge of the  
nuclear equation-of-state (EOS) at supra-normal densities and extreme 
isospin is essential for  
our understanding of the nuclear forces as well as for astrophysical  
purposes, the determination of the EOS was already one of the primary  
goals when first relativistic heavy ion beams started to operate in  
the beginning of the 80ties \cite{bevelac84}. In the following we will  
briefly discuss the knowledge on the nuclear EOS from a theoretical point of  
view, then turn to the realization within transport models,  
and finally give a short review on possible observables  
from heavy ion reactions to constrain the EOS. 
\section{Models for the nuclear EOS} 
Models which make predictions on the nuclear EOS can roughly be divided  
into three classes: 
\begin{enumerate} 
\item {\bf Phenomenological density functionals}: These are models based on  
effective density dependent interactions such as  
Gogny \cite{gogny,gogny02} or Skyrme forces \cite{skyrme04,reinhard04} or  
relativistic mean field (RMF) models \cite{rmf}.  
The number of parameters which are fine tuned to the nuclear chart is  
usually larger than six and less than 15. This type of models  
allows the most precise description of finite nuclei  
properties. 
\item {\bf Effective field theory approaches}: Models where  
the effective interactions is determined within the spirit of  
effective field theory (EFT) became recently more and more popular.  
Such approaches lead to a more systematic expansion of the  
EOS in powers of density, respectively the Fermi momentum $k_F$. They  
can be based on density functional theory \cite{eft1,eft2} or e.g. on  
chiral perturbation theory \cite{lutz00,finelli,weise04}. The advantage of EFT  
is the small number of free parameters and a correspondingly higher  
predictive power. However, when high precision fits to finite nuclei  
are intended this is presently only possible by the price of  
fine tuning through additional parameters. Then EFT functionals are  
based on approximately the same number of model parameters as   
phenomenological density functionals.   
\item {\bf Ab initio approaches}: Based on high precision  
free space nucleon-nucleon interactions, the nuclear many-body  
problem is treated microscopically. Predictions for the nuclear EOS  
are parameter free. Examples are variational  
calculations \cite{wiringa79,akmal98}, Brueckner-Hartree-Fock (BHF)  
\cite{gammel,jaminon,zuo02,zuo04}  
or relativistic Dirac-Brueckner-Hartree-Fock (DBHF)  
\cite{terhaar87,bm90,dejong98,boelting99,schiller01,honnef,dalen04}  
calculations and Greens functions  
Monte-Carlo approaches \cite{muether00,dickhoff04,carlson03}.  
\end{enumerate} 
Phenomenological models as well as EFT contain parameters which have  
to be fixed by nuclear properties around or below saturation density   
 which makes the extrapolation to supra-normal densities somewhat  
questionable. However, in the EFT case such an extrapolation is safer  
due to a systematic density  
expansion. One has nevertheless, to keep in mind that EFT approaches are  
based on low density expansions. Many-body calculations,  
on the other hand,  
have to rely on the summation of relevant diagram classes and are still  
too involved for systematic applications to finite nuclei.  
\subsection{Mean field theory} 
Among non-relativistic density functionals are Skyrme functionals  
most frequently used. The Skyrme interaction contains an  
attractive local two-body part and a repulsive density dependent  
two-body interaction which can be motivated by local three-body  
forces. We will not consider surface terms which involve gradients  
as well as spin-orbit contributions since they vanish in infinite  
nuclear matter.  
For a detailed discussion of Skyrme functionals and their  
relation to relativistic mean field (RMF) theory see e.g. \cite{reinhard04}.  
The EOS of symmetric nuclear matter, i.e. the binding energy per particle has the simple form 
\begin{equation} 
E/A = \frac{3 k_{F}^2}{10 M}  
+ \frac{\alpha}{2}\rho + \frac{\beta}{1+\gamma} \rho^\gamma~~,  
\label{Skyrme} 
\end{equation} 
where the first term in (\ref{Skyrme}) represents the kinetic energy  
of a non-relativistic Fermi gas and the remaining part the potential energy. 
To examine the structure of relativistic mean field models  
it is instructive to consider the simplest version of a relativistic  
model, i.e. the $\sigma\omega$ model of Quantum Hadron Dynamics  
(QHD-I) \cite{sw86}. In  QHD-I the  
nucleon-nucleon interaction is mediated by the exchange of  
two effective boson fields which  
are attributed to a scalar $\sigma$ and a vector $\omega$ meson.  
The energy density in infinite cold and isospin saturated nuclear  
matter is in mean field approximation given by  
\begin{equation} 
\epsilon = \frac{3}{4} E_F \varrho 
       + \frac{1}{4} m^{*}_D \,\varrho_{S} +  
\frac{1}{2}\left\{ \Gamma_V \,\varrho^2  
       + \Gamma_S\, \varrho_{S}^2 \right\}~~, 
\label{eqhd1} 
\end{equation} 
where the Fermi energy is given by $E_F =\sqrt{k_{F}^2 +  m^{*2}_D}$. 
We will denote  $m^{*}_D$ explicitely  
as Dirac mass in the following  
in order to distinguish it from its non-relativistic counterpart.   
The effective mass absorbs the scalar part of the mean field  
$m^{*}_D = M-\Gamma_S \varrho_{S}$. In the limit $ m^{*}_D\longrightarrow M$  
the first two terms in (\ref{eqhd1}) provide the energy  
(kinetic plus rest mass) of a non-interacting relativistic Fermi gas.  
 
A genuine feature of all relativistic  
models is the fact that one has to distinguish between the vector  
density $\varrho = 2k_{F}^3/3\pi^2$  
and a scalar density $\varrho_{S}$. The vector  
density is the time-like component of a 4-vector current $j_\mu$  
which spatial components vanish in the nuclear matter rest frame,  
while  $\varrho_{S}$ is a Lorentz scalar. The scalar density shows  
a saturation behavior with increasing vector density  
which is essential for the relativistic saturation mechanism.  
This becomes clear when binding energy $E/A = \epsilon/\varrho -M$   
is expanded in powers  of the Fermi momentum $k_F$ 
\beqa
E/A &=& \bigg[ \frac{3 k_{F}^2}{10 M} 
            - \frac{3 k_{F}^4}{56 M^3} 
            + \cdots \bigg] 
+\frac{1}{2} \bigg[\Gamma_V - \Gamma_S \bigg]\varrho  
\label{expand1}\\
&+&\Gamma_S \frac{\varrho}{M}\bigg[ \frac{3 k_{F}^2}{10 M} -   
\frac{36 k_{F}^2}{175 M^3} + \cdots \bigg]  
+ {\cal O}\left( (\Gamma_S \varrho/M)^2\right) 
\nonumber
\eeqa 
The first term in (\ref{expand1}) contains the kinetic energy  
of a non-relativistic Fermi gas followed by relativistic corrections and  
the remaining terms are the contributions from the mean field.  
In QHD-I the scalar and vector field strengths are given by the  
coupling constants for the corresponding mesons 
$ \Gamma_S = g_{\sigma}^2/m_{\sigma}^2$ and  
$\Gamma_V = g_{\omega}^2/m_{\omega}^2$ divided by the meson masses.  
The two parameters $ \Gamma_{S,V}$ are now fitted to the saturation  
point of nuclear matter $E/A\simeq  -16~ {\rm MeV}, 
~\varrho_{0} \simeq 0.16 ~{\rm fm}^{-3}$ which follows from the volume   
part of the Weiz\"acker mass formula.  
 The saturation mechanism requires that both coupling constants are  
large. This  leads automatically to the cancellation of two large  
fields, namely  
an attractive scalar field $\Sigma_S=-\Gamma_S \varrho_{S}$ and a repulsive  
vector field $\Sigma_V = \Gamma_V \varrho$. As is a typical feature  
of relativistic dynamics the single particle potential  
$U=m^{*}_D/E^*\Sigma_S - \Sigma_V ~ (E^* =\sqrt{ {\bf k}^2 +m^{*2}_D}) $,  
which is of the order of -50 MeV,  
results from the cancellation of scalar and vector fields, each of the order  
of several hundred MeV.  
 
However, with only two parameters QHD-I provides a relatively poor  
description of the saturation point with a too large saturation density  
and a very stiff EOS (K=540 MeV). To improve on this higher order corrections  
in density have to be taken into account. This can  
be done in several ways: In the spirit of the original Walecka model  
non-linear meson self-interaction terms have been introduced  
into the QHD Lagrangian \cite{rmf,bog82}. An alternative are  
relativistic point coupling  
models where the explicit meson exchange picture is abandoned. A  
Lagrangian of nucleon and boson fields with point couplings can be constructed  
in the spirit of EFT and expanded in powers of density  \cite{eft1,eft2}.  
Finite range effects from meson propagators are replaced by  
density gradients \cite{eft1,eft2}.  A third possibility is density  
dependent hadron field theory DDRH \cite{fule95,lenske00}. In DDRH the  
scalar and vector coupling constants are replaced by density  
dependent vertex functions $\Gamma_{S,V} (k_F)$. The density dependence  
of these renormalized vertices can either be taken from  
 Brueckner calculations thus parameterizing many-body correlations 
\cite{fule95,lenske00} or be determined  
phenomenologically \cite{typel99,niksic02}. In all  cases additional  
parameters are introduced which allow  a description of finite  
nuclei with a precision comparable to the best fits from Skyrme  
functionals. Phenomenological density functionals  
provide high quality fits to the known areas  
of the nuclear chart. Binding energies and rms-radii are reproduced  
with an average relative error of about $\sim 1-5$ \%. However, when the  
various models are extrapolated to the unknown regions of extreme  
isospin or to super-heavies predictions start to deviate substantially.  
This demonstrates the limited predictive power of these functionals. 
\subsection{Effective field theory} 
When concepts of effective field theory are applied to nuclear physics  
problem one has to rely on a separation of scales. EFT is based on a  
perturbative expansion of the nucleon-nucleon (NN) interaction or the  
nuclear mean field within  power counting schemes. The short-range  
part of the NN interaction requires a non-perturbative  
treatment, e.g. within the Brueckner ladder summation. The philosophy  
behind EFT is to separate short-range correlations  
from the long and intermediate range part of the NN-interaction.  
This assumption is motivated by the fact that the scale of the  
short-range correlations, i.e.  the hard core, is set by the  
$\rho$ and $\omega$ vector mesons masses which lie well above  
the Fermi momentum and the pion mass which sets the scale of the 
long range forces.  
\begin{figure}[h] 
\resizebox{0.75\columnwidth}{!}{\includegraphics{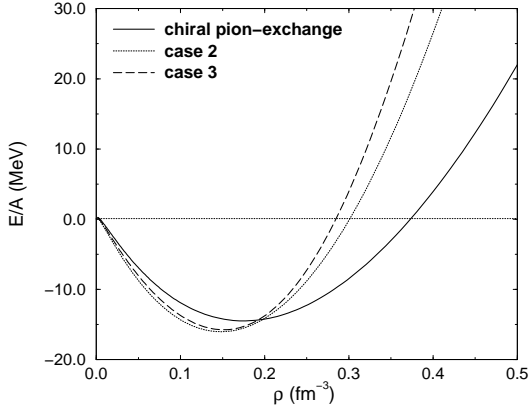}}
\caption{EOS for symmetric nuclear  
matter obtained from chiral one- and two-pion exchange (case 1, solid curve),  
by adding background fields from QCD sum rules, and  
finally after fine tuning to finite nuclei properties  
(case 3, dashed curve). Figure is taken from \protect\cite{finelli}. 
} 
\label{eft1_fig} 
\end{figure} 
The density functional theory (DFT) formulation of the  
relativistic nuclear many-body problem \cite{eft1,eft2} is thereby  
analogous to the Kohn--Sham approach in DFT.  
An energy functional of scalar and vector densities is constructed  
which by minimization gives rise to variational equations that 
determine the ground-state densities.  
Doing so, one tries to approximate the {\it exact} functional  
using an expansion in classical meson fields and their derivatives,  
based  
on the observation that the ratios of these quantities to the nucleon mass 
are small, at least up to moderate density.  
The exact energy functional which one tries to derive  
explicitely when using many-body techniques  
such as Brueckner or variational approaches contains  
exchange-correlation and all other many-body and  
relativistic effects. The DFT interpretation  
implies that the model parameters fitted to nuclei 
implicitly contain effects of both short-distance physics and many-body 
corrections. 
 
Recently also concepts of chiral perturbation theory (ChPT)  
have been applied to the nuclear many-body problem  
\cite{finelli,weise04}. Doing so, the long  
and intermediate-range interactions are treated explicitely within  
chiral pion-nucleon dynamics. This allows an expansion of the  
energy density functional in powers of  $m_\pi/M$  
or in $k_F/M$. Like in DFT,  
short-range correlation are not resolved explicitely  
but handled by counter-terms (dimensional regularization)  
\cite{lutz00} or through a cut-off  
regularization \cite{weise04}. Fig. \ref{eft1_fig} shows  
the corresponding EOS obtained from  
chiral one- and two-pion exchange between nucleons. In  
order to account for the most striking feature  
of relativistic dynamics, expressed by the existence of  
the large scalar and vector fields, in Refs. \cite{finelli,weise04}  
 iso-scalar condensate background nucleon self-energies derived  
from QCD sum rules have been added to the chiral fluctuations.  
To lowest order in density the QCD condensates give rise to  
a scalar  self-energy  
$\Sigma_S = - \sigma_N M/(m_\pi^2 f_\pi^2) \varrho_S $ and  
a vector  self-energy  
$\Sigma_V = 4 (m_u + m_d)M/(m_\pi^2f_\pi^2) \varrho$.  
It is remarkable that the total self-energies, i.e.  
condensates plus chiral fluctuations, are very close to those  
obtained from DBHF calculations \cite{finelli,boelting99}.  
The resulting EOS is also shown in  Fig. \ref{eft1_fig} in addition  
with that obtained after fine tuning to finite  
nuclei. Although the original EOS (case 1) is rather soft the  
inclusion of the condensates and the adjustment to finite nuclei  
results in an EOS with is finally stiff. 
\subsection{Ab initio calculations} 
\begin{figure}[h] 
\resizebox{0.9\columnwidth}{!}{\includegraphics{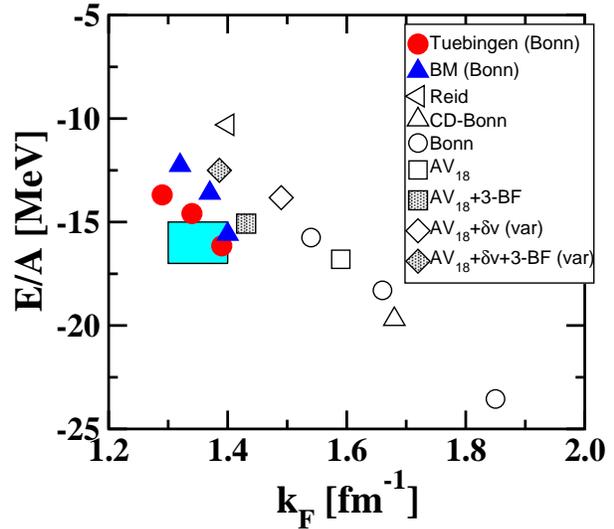}}
\caption{Nuclear matter saturation points from relativistic (full symbols)  
and non-relativistic (open symbols) Brueckner-Hartree-Fock calculations  
based on different nucleon-nucleon forces. The diamonds show results from  
variational calculations. Shaded symbols denote calculations which  
include 3-body forces. The shaded area is the empirical region of  
saturation. 
} 
\label{coester_fig} 
\end{figure} 
In  {\it ab initio} calculations based on many-body techniques one  
derives the energy functional from first principles, i.e. treating  
short-range and  many-body correlations explicitely. A typical  
example for a successful  many-body approach is Brueckner theory \cite{gammel}.  
In the relativistic Brueckner approach the nucleon  
inside the medium is dressed by the self-energy $\Sigma$.  
The in-medium T-matrix which is obtained from  
the relativistic Bethe-Salpeter (BS) equation plays the role  
of an effective two-body interaction which contains all short-range  
and many-body correlations of the ladder approximation.  
Solving the BS-equation the Pauli principle is respected  
and intermediate scattering states are projected  
out of the Fermi sea.  
The summation of the  T-matrix over the occupied states inside the Fermi sea  
yields finally the self-energy in Hartree-Fock approximation. This coupled set of  
equations states a self-consistency problem which has to be solved  
by iteration.  
 
In contrast to relativistic DBHF calculations which came up in the late 
80ties non-relativistic BHF theory has already almost half a century's 
history. The first numerical calculations for nuclear matter were carried 
out by Brueckner and Gammel in 1958 \cite{gammel}. Despite strong  
efforts invested in the development of improved solution techniques for  
the Bethe-Goldstone (BG) equation, the non-relativistic counterpart of the  
BS equation, it turned out that, although such calculations were able to 
describe the nuclear saturation mechanism qualitatively, they failed  
quantitatively. Systematic studies for a large number of NN 
interactions were always allocated on a 
so-called {\it Coester-line} in the $E/A-\rho$ plane which does not 
meet the empirical region of saturation. In particular modern  
one-boson-exchange (OBE) potentials  
lead to strong over-binding and too large saturation densities where  
relativistic calculations do a much better job.  
 
Fig. \ref{coester_fig} compares the saturation points of nuclear matter  
obtained by relativistic Dirac-Brueckner-Hartree-Fock (DBHF) calculations  
using the  Bonn potentials \cite{bonn} as bare $NN$ interactions   
to non-relativistic Brueckner-Hartree-Fock calculations for various  
 $NN$ interactions. The DBHF results are taken from Ref. \cite{bm90} (BM)  
and more recent calculations based on improved techniques are  
from \cite{boelting99} (T\"ubingen). 
\begin{figure}[h] 
\resizebox{0.9\columnwidth}{!}{\includegraphics{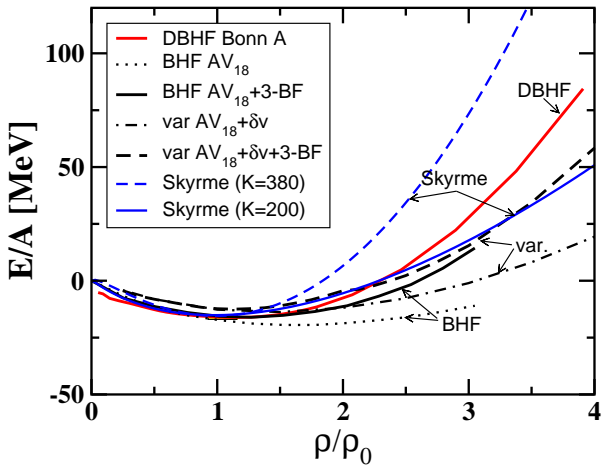}}
\caption{Predictions for the EOS of symmetric  
nuclear matter from microscopic ab initio  
calculations, i.e. relativistic DBHF \protect\cite{boelting99},  
non-relativistic BHF \protect\cite{zuo02} and variational  
\protect\cite{akmal98} calculations. For comparison also  
 soft and hard Skyrme forces are shown. 
}
\label{dbeos_fig} 
\end{figure} 
Several reasons have been discussed in the literature in order to 
explain the success of the relativistic treatment.  
The saturation mechanisms in relativistic and non-relativistic 
theories are quite different. In relativistic MFT the vector field grows  
linear with density while  
the scalar field saturates at large densities.  
The magnitude and the density dependence of the 
scalar and vector DBHF self-energy is similar to MFT, i.e. the  
single particle potential is the result of the cancellation of two  
large scalar and vector fields, each several hundred MeV in magnitude  
(see e.g. the effective mass in Fig. \ref{mass1_fig}). In BHF,  
on the other hand,  the  saturation mechanism takes place  
exclusively on the scale of the binding energy, i.e. a few ten MeV. It  
cannot be understood by the absence of a tensor 
force. In particular the second order 1-$\pi$ exchange potential  
(OPEP) is large and 
attractive at high densities and its interplay with Pauli-blocking 
leads finally to saturation. Relativistically the tensor force  
is quenched by a factor $(m^{*}_D/M)^2$ and less important  
for the saturation mechanism \cite{banerjee02}. 
 
Three-body forces (3-BFs) have extensively been studied within 
non-relativistic BHF \cite{zuo02} and variational calculations  
\cite{akmal98}. The contributions from 3-BFs are in total   
repulsive which makes the EOS harder and non-relativistic  
calculations come close to their relativistic counterparts.  
The same effect is observed in variational calculations \cite{akmal98}  
shown in Fig. \ref{dbeos_fig}. The variational results shown  
contain boost corrections ($\delta v$) which account for relativistic kinematics and 
lead to additional repulsion \cite{akmal98}. Both, BHF 
\cite{zuo02} and the variational calculations from  
 \cite{akmal98} are based on the latest ${\rm AV}_{18}$ version of the Argonne  
potential. In both cases phenomenological 3-body-forces are used,   
the Tucson-Melbourne 3-BF in  \cite{zuo02} and  
the Urbana IX 3-BF\footnote{Using boost corrections  
the repulsive contributions of the UIX interaction are  
reduced by about 40\% compared to the original ones in \cite{akmal98}} 
in \cite{akmal98}.  
It is often argued that in non-relativistic 
treatments 3-BFs play in some sense an equivalent role as the 
dressing of the two-body interaction by in-medium spinors in  
Dirac phenomenology. Both mechanisms lead indeed to an effective density 
dependent two-body interaction $V$ which is, however, of different 
origin. One class of 3-BFs involves virtual excitations of 
nucleon-antinucleon pairs. Such Z-graphs  are 
in net repulsive and can be considered as a renormalization of  
the meson vertices and propagators. A second class of 3-BFs is  
related to the inclusion of explicit resonance degrees of freedom.  
The most important resonance is the $\Delta$(1232)  
isobar which provides at low and intermediate  
energies large part of the intermediate range attraction.  
Intermediate $\Delta$ states appear in 
elastic $NN$ scattering only in combination with at least two-isovector-meson 
exchange ($\pi\pi,~\pi\rho,\dots$). Such box diagrams  
can satisfactorily be absorbed into an effective 
$\sigma$-exchange \cite{bonn}. The maintenence of  
explicit $\Delta$ DoFs gives  rise to  
additional saturation, shifting the saturation point away  
from the empirical region \cite{terhaar87}. However,  
as pointed out e.g. in \cite{muether00} the 
inclusion of non-nucleonic DoFs has to be 
performed with caution: Freezing out resonance DoFs  
generates automatically a class of three-body forces which contains  
nucleon-resonance excitations. There exist strong 
cancellation effects between the repulsion due to box diagrams    
and contributions from 3-BFs. Non-nucleonic DoFs 
and many-body forces should therefore be treated on the same footing.  
Such a treatment may be possible with the next generation of  
nucleon-nucleon forces based on  chiral perturbation theory  
\cite{klock94,entem03} which allows a systematic  
generation of three-body forces.  Next-to-leading order all 3-BFs cancel  
while non-vanishing contributions appear at NNLO.

Fig. \ref{dbeos_fig} compares the equations of state from the  
different approaches: DBHF from Ref. \cite{boelting99}   
based the Bonn A interaction\footnote{The high density behavior of the  
EOS obtained with different interaction, e.g. Bonn B or C is  
very similar. \cite{boelting99}}  \cite{bonn},  
BHF  \cite{zuo02}  
and variational calculations \cite{akmal98}. The latter ones are  
 based on the Argonne ${\rm AV}_{18}$ potential and include  
3-body forces. All the approaches use modern high precision $NN$  
interactions and represent state of the art calculations. Two   
phenomenological Skyrme functionals which correspond to the  
limiting cases of a soft (K=200 MeV) and a hard (K=380 MeV) EOS are  
shown as well. In contrast to the Skyrme  
interaction (\ref{Skyrme})  
where the high density behavior is fixed by the compression modulus,  
in microscopic approaches the compression modulus is  
only loosely connected to the curvature at saturation density.   
DBHF Bonn A has e.g. a compressibility of K=230 MeV.  
Below $3\rho_0$ both are not too far  
from the soft Skyrme EOS. The same is true for BHF including 3-body  
forces.  

When many-body calculations are performed, one has to keep in mind that 
elastic $NN$ scattering data constrain the interaction only 
up to about 400 MeV, which corresponds to the pion 
threshold. $NN$ potentials differ essentially 
in the treatment of  the short-range part. A model independent 
representation of the $NN$ interaction can be obtained in EFT approaches where 
the unresolved short distance physics is replaced  by simple contact 
terms. In the framework of chiral EFT the $NN$ interaction has been 
computed up to N$^3$LO \cite{entem03,epelbaum05}. An alternative 
approach which leads to similar results is based on 
renormalization group (RG) methods \cite{lowk03}. In the $V_{\rm low~k}$ 
approach a low-momentum potential is derived from a given 
realistic $NN$ potential by integrating out 
the high-momentum modes using RG methods. 
At a cutoff $\Lambda \sim 2~{\rm fm}^{-1}$ all the different $NN$ potential models
were found to collapse to a model-independent effective 
interaction $V_{\rm low~ k}$. When applied to the nuclear many-body 
problem low momentum interactions do not require a full resummation 
of the Brueckner ladder diagrams but can already 
be treated within second-order perturbation theory \cite{bogner05}. 
However, without  repulsive three-body-forces 
isospin saturated nuclear matter was found to collapse. Including 3-BFs 
first promising results have been obtained with $V_{\rm low~ k}$ 
\cite{bogner05}, however, nuclear saturation is not yet described 
quantitativley. Moreover, one has to keep in mind that - 
due to the high momentum cut-offs - EFT is essentially 
suitable at moderate densities. 
\section{EOS in symmetric and asymmetric nuclear matter} 
\begin{figure}[h] 
\resizebox{0.99\columnwidth}{!}{\includegraphics{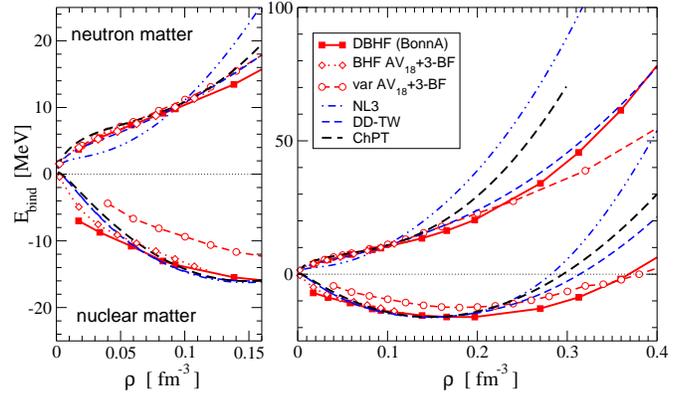}}
\caption{EOS in nuclear matter and neutron matter.  
BHF/DBHF and variational calculations are compared to  
phenomenological density functionals NL3 and DD-TW and  
ChPT+corr.. The left panel zooms the low density range.  
} 
\label{nmeos_fig} 
\end{figure} 
Fig. \ref{nmeos_fig} compares now the predictions for nuclear and neutron  
matter from microscopic  
many-body calculations -- DBHF \cite{dalen04}  and the 'best' 
variational calculation with 3-BFs and boost corrections  
\cite{akmal98} -- to phenomenological approaches and to EFT. As typical  
examples for relativistic functionals we take NL3 \cite{nl3} as one  
of the best RMF fits to the nuclear chart and a  
phenomenological density dependent  
RMF functional DD-TW from \cite{typel99}.  
ChPT+corr. is based on chiral pion-nucleon dynamics  
including condensate fields and fine tuning to finite nuclei (case 3 in  
Fig. \ref{eft1_fig}). As expected the phenomenological functionals  
agree well at and below saturation density where they are constrained  
by finite nuclei, but start to deviate substantially at supra-normal  
densities. In neutron matter the situation is even worse since  
the isospin dependence of the phenomenological functionals is less constrained.  
The predictive power of such density functionals  at supra-normal  
densities is restricted.   {\it Ab initio}  
calculations predict throughout a soft EOS in the density range  
relevant for heavy ion reactions at intermediate and low  
energies, i.e. up to about three times $\rho_0$.  
There seems to be no way to obtain an  
EOS as stiff as the hard Skyrme force shown in  Fig. \ref{dbeos_fig} or NL3. 
Since the $nn$ scattering lenght is large, neutron matter 
at subnuclear densities is less  
model dependent. The microscopic 
calculations (BHF/DBHF, 
variational) agree well and results are consistent with 
 'exact' Quantum-Monte-Carlo calculations \cite{carlson03}. 
\begin{figure}[h] 
 \resizebox{0.98\columnwidth}{!}{\includegraphics{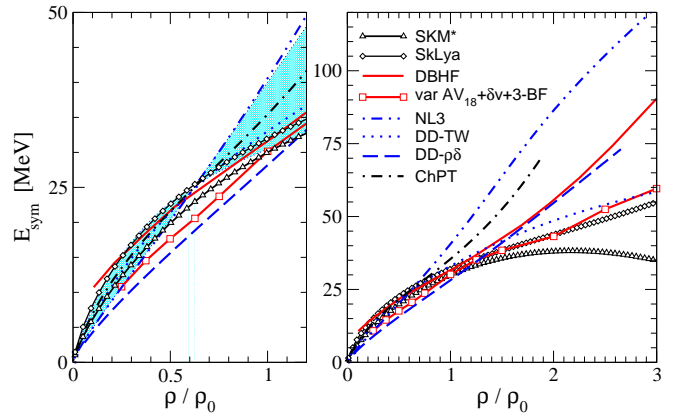}}
\caption{Symmetry energy as a function of density as predicted by different  
models. The left panel shows the low density region while the right  
panel displays the high density range. 
} 
\label{esym_fig} 
\end{figure} 

In isospin asymmetric matter the binding energy is a functional  
of the proton and neutron densities, 
characterized by the  asymmetry parameter $\beta=Y_n-Y_p$ which  
is the difference of the neutron and proton fraction  
$Y_i=\rho_i/\rho~,i=n,p$.  The isospin dependence of the energy   
functional can be expanded in terms of $\beta$ which leads to a  
parabolic dependence on the asymmetry parameter 
\beqa 
E(\rho,\beta) &=& E(\rho) + E_{\rm sym}(\rho) \beta^2 + {\cal O}(\beta^4) +  
\cdots \nonumber \\ 
E_{\rm sym}(\rho) &=& \frac{1}{2}   
\frac{\partial^2E(\rho,\beta)}{\partial \beta^2}|_{\beta=0}  
= a_4 + \frac{p_0}{\rho_{0}^2} (\rho - \rho_0) +\cdots  
\label{esym} 
\eeqa
Fig. \ref{esym_fig} compares the symmetry energy predicted from  
the DBHF and variational calculations to that of the  
empirical density functionals already shown in Fig. \ref{nmeos_fig} 
In addition the relativistic DD-$\rho\delta$ RMF functional \cite{baran04}  
is included.  
Two Skyrme functionals, SkM$^*$ and the more recent Skyrme-Lyon force  
SkLya represent non-relativistic models.   
The left panel zooms the low density region while the right panel  
shows the high density behavior of $ E_{\rm sym}$. Remarkable  
is that most empirical models coincide around $\rho \simeq 0.6 \rho_0$  
where   $ E_{\rm sym} \simeq 24$ MeV. This demonstrates that  
constraints from finite nuclei are active for an average density  
slightly above half saturation density. 
However, the extrapolations to supra-normal densities  
diverge dramatically. This is crucial since the high density behavior  
of  $ E_{\rm sym}$ is essential for the structure and the stability of  
neutron stars (see also the discussion in Sec. V.5).  
The microscopic models show a density dependence which can still  
be considered as {\it asy-stiff}. DBHF \cite{dalen04} is thereby stiffer  
than the variational results of \cite{akmal98}. The density  
 dependence is generally more complex  than in RMF  theory, in  
particular at high densities where $E_{\rm sym}$ shows a non-linear and  
more pronounced increase.   
Fig. \ref{esym_fig} clearly demonstrates the necessity to constrain  
the symmetry energy at supra-normal densities with the help of heavy  
ion reactions. 
\begin{figure}[h] 
\resizebox{0.75\columnwidth}{!}{\includegraphics{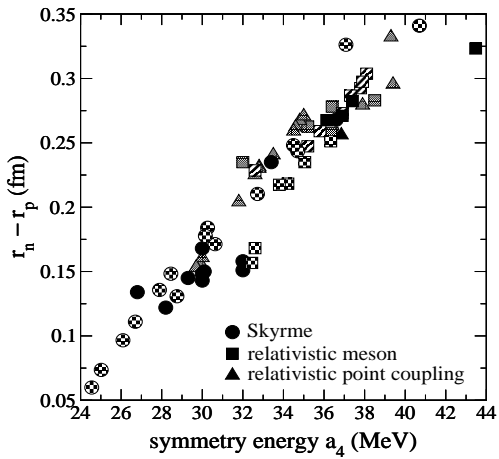}}
\caption{Skin thickness in $^{208}$Pb versus the linear symmetry energy  
parameter $a_4$ for various models. Figure is taken from  
\protect\cite{eft2}.  
} 
\label{esym2_fig} 
\end{figure} 
The hatched area in Fig. \ref{esym_fig}  
displays the range of $ E_{\rm sym}$ which has  
been obtained by constructing a density dependent  
RMF functional varying thereby the  
 linear asymmetry parameter $a_4$ from 30 to 38 MeV \cite{niksic02}.  
In  \cite{niksic02} it was concluded that  
charge radii, in particular the skin thickness $r_n - r_p$ in heavy  
nuclei constrains the allowed  
range of  $a_4$ to $32\div 36$ MeV for relativistic functionals. 
 
Fig. \ref{esym2_fig} displays the correlation between the  
 skin thickness in $^{208}$Pb and $a_4$ obtained within various models.  
The skin thickness depends, however, not only on the symmetry energy  
but there exists a close correlation between $a_4$ and the  
compression modulus K \cite{niksic02}. This correlation is of  
importance when these quantities are extracted from finite nuclei  
(see discussion by Shlomo et al. in Sec. I.4). 
\subsection{Effective nucleon masses} 
The introduction of an effective mass is a common concept to characterize the  
quasi-particle properties of a particle inside a strongly interacting  
medium. In nuclear physics exist different definitions of the  
effective nucleon mass which are  
often compared and sometimes even mixed up:  
the non-relativistic effective mass $m^*_{NR}$ and the  
relativistic Dirac mass $m^*_{D}$. 
These two definitions  are based on different physical concepts.  
The nonrelativistic mass parameterizes 
the momentum dependence of the single-particle potential.  
The relativistic Dirac mass  
is defined through the scalar part of the nucleon self-energy in the  
Dirac field equation which is absorbed into the effective mass  
$m^*_{D} =M + \Sigma_S (k, k_F)$. The Dirac mass is a smooth function of  
the momentum. In contrast, the nonrelativistic effective  
mass - as a model independent result -   
shows a narrow enhancement near the Fermi surface due to an enhanced  
level density \cite{mahaux85}. For a recent review on this subject 
and experimental constraints on  $m^*_{NR}$ see \cite{lunney03}. 

While the Dirac  mass is a genuine relativistic quantity the effective  
mass  $m^*_{NR}$ is determined by the single-particle energy  
\begin{eqnarray} 
m^*_{NR} = k [dE/dk]^{-1} = \left[\frac{1}{M}  
+  \frac{1}{k} \frac{d}{ dk} U \right]^{-1}~~. 
\label{Landau1} 
\end{eqnarray} 
\begin{figure}[h] 
\resizebox{0.98\columnwidth}{!}{\includegraphics{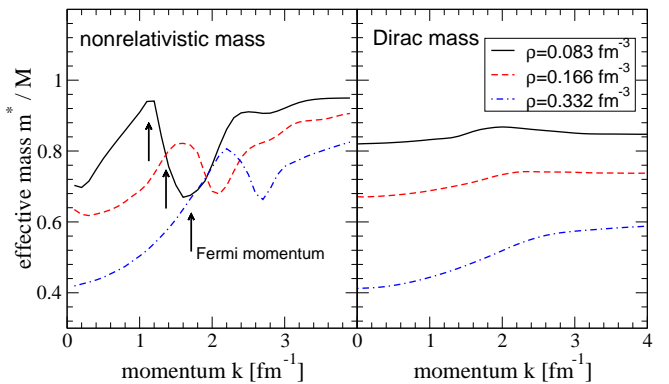}}
\caption{The effective mass in isospin symmetric nuclear matter  
as a function of the momentum $k$ at different densities  
determined from relativistic Brueckner calculations.   
} 
\label{mass1_fig} 
\end{figure} 
$m^*_{NR}$ is a measure of the non-locality of  
the single-particle potential $U$ (real part) which can be due to  
non-localities in space, resulting in  
a momentum dependence, or in time, resulting in an energy dependence.  
In order to clearly separate both effects, one has to distinguish further  
between the so-called k-mass and the E-mass \cite{jaminon}.  
The spatial non-localities of $U$ are mainly  
generated by exchange Fock terms and the resulting k-mass is a smooth  
function of the momentum. Non-localities in time are generated by Brueckner  
ladder correlations due to the scattering to intermediate states which  
are off-shell. These are mainly short-range correlations which generate a  
strong momentum  dependence with a characteristic enhancement of the  
E-mass slightly above the Fermi surface \cite{mahaux85,jaminon,muether04}.  
The effective mass defined  
by Eq. (\ref{Landau1}) contains both, non-localities in  
space and time and is given by the product of k-mass and E-mass \cite{jaminon}.  
In Fig.~\ref{mass1_fig}  
the nonrelativistic effective mass and the Dirac mass, both determined from  
DBHF calculations \cite{dalen05}, are  
shown as a function of momentum $k$ at different Fermi momenta  
of $k_F=1.07,~1.35,~1.7~{\rm fm}^{-1}$.  
$m^*_{NR}$  shows  the  typical peak structure as a function of  
momentum around $k_F$ which is also seen in BHF calculations \cite{muether04}.  
The peak reflects the increase of the  
level density due to the vanishing imaginary part of the optical  
potential at $k_F$ which is also seen, e.g., in shell model calculations  
\cite{mahaux85,jaminon}. One has, however,  
to account for correlations beyond mean field or Hartree-Fock  
in order to reproduce this behavior.  
\begin{figure}[h] 
\resizebox{0.98\columnwidth}{!}{\includegraphics{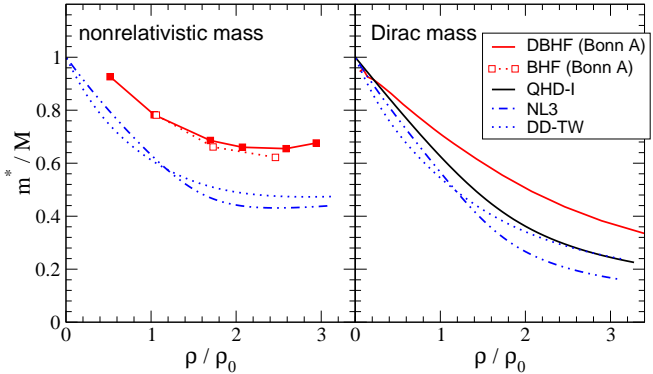}}
\caption{Nonrelativistic and Dirac effective  
mass in isospin symmetric nuclear matter  
as a function of the density for various models. 
} 
\label{mass2_fig} 
\end{figure} 
Fig.~\ref{mass2_fig} compares the density dependence of the two  
effective masses determined at $k_F$. Both masses decrease with increasing density,  
the  Dirac mass continously, while $m^*_{NR}$ starts to rise  
again at higher densities. Phenomenological density functionals  
(QHD-I, NL3, DD-TW) yield systematically smaller values of  $m^*_{NR}$  
than the microscopic approaches. This reflects the  
lack of nonlocal contributions from short-range and many-body  
correlations in the mean field approaches. 
\subsubsection{Proton-neutron mass splitting} 
A heavily discussed topic is in the moment the proton-neutron mass  
splitting in isospin asymmetric nuclear matter. This question is of  
importance for the forthcoming new generation of radioactive beam  
facilities which are devoted to the investigation of the isospin  
dependence of the nuclear forces at its extremes. However, presently  
the predictions for the  
isospin dependences differ substantially.  
BHF calculations \cite{zuo02,muether04}  
predict a proton-neutron mass splitting of $m^*_{NR,n} > m^*_{NR,p}$.  
This stands in contrast to relativistic mean-field (RMF) theory. When only a  
vector isovector $\rho$-meson is included Dirac phenomenology  
predicts equal masses $m^*_{D,n}= m^*_{D,p}$ while the inclusion of the  
scalar isovector $\delta$-meson, i.e. $\rho+\delta$, leads to  
$m^*_{D,n} < m^*_{D,p}$ \cite{baran04}. When the effective mass is  
derived from RMF theory, it shows the same behavior as the corresponding  
Dirac mass, namely  $m^*_{NR,n} < m^*_{NR,p}$ \cite{baran04}. Conventional  
Skyrme forces, e.g. SkM$^*$, lead to $m^*_{NR,n} < m^*_{NR,p}$ \cite{pearson01} 
while  the more recent Skyrme-Lyon interactions (SkLya) predict the same  
mass splitting as RMF theory.     
The predictions from relativistic DBHF calculations are in the  
literature still controversial. They depend strongly  on approximation schemes and  
techniques used to determine the Lorentz  
and the isovector structure of the nucleon self-energy.  
In the approach originally proposed by Brockmann and Machleidt  \cite{bm90}  
one extracts the scalar and vector  
self-energy components directly from the single-particle potential.  
Thus, by a fit to the single-particle potential  
mean values for the self-energy components are obtained where   
the explicit momentum-dependence has already been averaged out. In  
symmetric nuclear matter this method is relatively reliable but the  
extrapolation to asymmetric matter is ambiguous  \cite{schiller01}.   
Calculations based on this method  
predict a mass splitting of $m^*_{D,n} > m^*_{D,p}$~\cite{alonso03}.  
On the other hand, the components of the self-energies can directly  
be determined from the projection onto Lorentz invariant  
amplitudes \cite{terhaar87,dejong98,boelting99,schiller01,dalen04,horowitz87}.  
Projection techniques are involved but more accurate and  
yield the same mass splitting as found in RMF theory when the  
$\delta$ -meson is included, i.e. $m^*_{D,n} < m^*_{D,p}$  
\cite{dejong98,schiller01,dalen04}. Recently also the non-relativistic  
effective mass has been determined with the DBHF approach and here  
a reversed proton-neutron mass splitting was found, i.e. $m^*_{NR,n} >m^*_{NR,p}$  
\cite{dalen05}. Thus DBHF is  in agreement with the results from nonrelativistic  
BHF calculations. 

Experimentally accessable is the $p-n$ mass splitting, or 
the magnitude of the corresonding isovector effective mass $m^*_{V},~ 
(\frac{\beta}{m^*_{V}} =  \frac{\beta +1}{m^*_{NR}} - \frac{1}{m^*_{NR,n}})$ through 
the electric dipole photoabsorption cross section, i.e. through an enhancement 
of the Thomas-Reiche-Kuhn sum rule by the factor 
$m/m^*_{V}$. However, values derived from 
GDR measurements range presently from $m^*_{V}/m = 0.7\div 1.05$ 
\cite{lunney03,krivine80,khan02}. 
The forthcoming radioactive beam facilitites will certainly improve on this 
not yet satisfying situation.         
\subsection{Optical potentials} 
\begin{figure}[h] 
\resizebox{1.0\columnwidth}{!}{\includegraphics{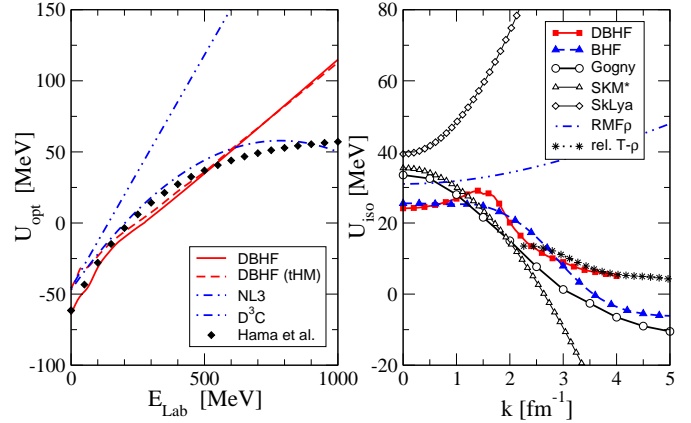}}
\caption{Nucleon optical potential in nuclear matter at $\rho_0$. 
On the left side DBHF calculations 
for symmetric nuclear matter from
\protect\cite{terhaar87} and \protect\cite{boelting99} are compared to the 
phenomenological models NL3 and D$^3$C \protect\cite{typel05} and to the 
p-A scattering analysis of \protect\cite{hama}. The right panel compares 
the iso-vector optical potential from DBHF \protect\cite{dalen04} and 
BHF   \protect\cite{zuo05} to phenomenological RMF 
\protect\cite{gaitanos04b} , Gogny and 
Skyrme forces and to a relativistic $T-\rho$ 
approximation \protect\cite{trho05}. 
} 
\label{uopt_fig} 
\end{figure}

The second important quantity related to the momentum dependence 
of the mean field is the optical nucleon-nucleus potential. At subnormal 
densities the optical potential $U_{\rm opt}$ is constraint by 
proton-nucleus scattering data \cite{hama} and at supra-normal densities 
constraints can be derived from heavy ion reactions 
\cite{dani00,gaitanos01,giessen2}. In a relativistic framework 
the optical Schroedinger-equivalent 
nucleon potential (real part) is defined as 
\begin{equation}
U_{\rm opt} 
=  - \Sigma_S  + \frac{E}{M} \Sigma_{V} 
        + \frac{\Sigma_S^2  - \Sigma_{V}^2}{2M}~.
\label{uopt}
\end{equation}
One should thereby note that in the literature sometimes also 
an optical potential, given by the difference of the single-particle 
energies in  medium and free space $U= E - \sqrt{M^2 +{\bf k}^2}$ is  
used \cite{dani00} which should be not mixed up with (\ref{uopt}). 
In a relativistic framework momentum independent fields $\Sigma_{S,V}$ 
(as e.g. in RMF theory) 
lead always to a linear energy dependence of $U_{\rm opt}$. 
As seen from Fig.~\ref{uopt_fig} DBHF reproduces 
the empirical optical potential \cite{hama} extracted from 
proton-nucleus scattering for nuclear matter at $\rho_0$ reasonably well 
up to a laboratory energy of about 0.6-0.8 GeV. 
However, the saturating behavior at large momenta cannot 
be reproduced by this calculations because of missing 
inelasticities, 
i.e. the excitation of isobar resonances above the pion threshold. When such continuum 
excitations are accounted for optical model caculations are 
able to describe nucleon-nucleus scattering data also at 
higher nergies \cite{geramb02}. In heavy ion 
reactions at incident energies above 1 AGeV such a saturating behavior is 
required in order to reproduce transverse flow observables \cite{giessen2}.
One has then to rely on phenomenological approaches where 
the strength of the vector potential is artificially suppressed, e.g. by 
the introduction of additional form factors \cite{giessen2} or by 
energy dependent terms in the QHD Lagrangian \cite{typel05} (D$^3$C model 
in Fig.\ref{uopt_fig}) . 

The isospin dependence, expressed by the 
isovector optical potential $U_{\rm iso}= (U_{\rm opt,n} - U_{\rm opt,p}) / (2
\beta)$ is much less constrained by data. The knowledge of 
this quantity is, however, 
of high importance for the forthcoming radioactive beam experiments.  
The right panel of Fig. \ref{uopt_fig} compares the predictions 
from DBHF \cite{dalen04} and BHF \cite{zuo05} to the phenomenological 
Gogny and Skyrme (SkM$^*$ and SkLya) forces and a relativistic $T-\rho$ 
approximation \cite{trho05} based on empirical NN scattering 
amplitudes \cite{neil83}. At large momenta 
DBHF agrees with the tree-level results of \cite{trho05}.
While the dependence of  $U_{\rm iso}$ on the 
asymmetry parameter $\beta$ is found to be rather weak \cite{dalen04,zuo05}, 
the predicted energy and density dependences are quite 
different,  in particular between  the microscopic and the 
phenomenological approaches. The energy dependence 
of $U_{\rm iso}$ is very little constrained by data. The old 
analysis of optical potentials of scattering on charge asymmetric 
targets by Lane \cite{lane62} is consistent with a decreasing 
potential as predicted by DBHF/BHF, while more recent analyses 
based on Dirac phenomenology \cite{madland89} come to the opposite 
conclusions. RMF models show a linearly increasing
 energy dependence of $U_{\rm iso}$ (i.e. quadratic in $k$)  
like SkLya, however 
generally with a smaller slope (see discussion in \cite{baran04}). 
To clarify this question certainly more experimental 
efforts are necessary.  
\section{Transport models} 
The difficulty to extract information on the EOS from 
heavy ion reactions lies in the fact that the colliding system is over 
a large time span of the reaction out of global and even local 
equilibrium. At intermediate energies the relaxation time needed 
to equilibrate coincides more or less with the high density phase of 
the reaction. Hence, non-equilibrium effects are present 
all over the compression phase where one essentially intends to 
study the EOS at supra-normal densities. 
Experimental evidences for incomplete equilibration 
even in central collisions have been found by isospin tracing 
of projectile and target nuclei \cite{rami99} and by different 
variances of longitudinal and transverse rapidity distributions 
\cite{reisdorf04}. To account for the 
temporal space time evolution of the reactions requires dynamical 
approaches which are based on kinetic transport theory. In the following 
we briefly discuss the various approaches which are mainly used in order 
to describe the reaction dynamics at low and intermediate energies.  
\subsection{Boltzmann-type kinetic equations} 
The theoretical basis for the  description of the collision dynamics at 
energies ranging from the Fermi regime up to 1-2 AGeV is hadronic 
non-equilibrium quantum transport field theory \cite{btm90}. 
The starting point of non-equilibrium QFT is 
the Schwinger-Keldysh formalism for many-body 
Green functions in non-equilibrium configurations. The one-body Green function  
is defined as the expectation value of the time ordered product of 
fermionic field operators  
$G(1,1') = (-i) <{\cal T}_{sk}( \Psi(1) \bar{\Psi}(1')) > $ where 
${\cal T}_{sk}$ defines the temporal sequence  
of the field operators. In non-equilibrium time reversal  
invariance is violated and thus the application of ${\cal T}_{sk}$ leads 
to four possible combinations \cite{btm90}: 
\begin{eqnarray} 
G^{c}  &=&  -i \langle T^{c} [ \Psi(1) \overline{\Psi}(1') ] \rangle,~~  
G^{a}   =  -i \langle T^{a} [ \Psi(1) \overline{\Psi}(1') ] \rangle,~~  
\nonumber\\ 
G^{>}  &=&  -i \langle \Psi(1) \overline{\Psi}(1') \rangle,~~ 
~~~~~G^{<}  =  i\langle  \overline{\Psi}(1') \Psi(1) \rangle  
\label{green_kontur} 
\end{eqnarray} 
where $T^{c}$ ($T^{a}$) is the causal (anti-causal) time ordering operator.  
The physical quantity of interest is the 
correlation function $G^{<}$ since it corresponds in the equal time limit  
to the  density 
$ \lim_{t_{1'} \to t_{1}}G^{<}(1,1') = (+i)\rho({\bf x}_{1},{\bf x}_{1'},t)$. 
However, the four Green-functions are related through equations of 
motion (Kadanoff-Baym Equations) for  
the correlation $G^{<,>}$ and the retarded and advanced $G^{\pm}$ functions  
(the retarded and advanced Green functions are defined via  
$G^{+,-}=G^{c}-G^{<,>}=G^{>,<}-G^{a}$). 
From the Kadanoff-Baym Eqs. one obtains a  
kinetic equation for the correlation function $G^{<}$ 
\begin{eqnarray} 
&&D G^{<} - G^{<} D^{*}  
- \left( Re \Sigma^{+} G^{<} - G^{<} Re \Sigma^{+} \right)  
- \left( \Sigma^{<} Re G^{+} \right.
\nonumber \\
&& \left. - Re G^{+} \Sigma^{<} \right)  
 = \frac{1}{2}  
  \left(  
	\Sigma^{>} G^{<} + G^{<} \Sigma^{>} - \Sigma^{<} G^{>} - G^{>} \Sigma^{<}  
  \right)  
\label{TP1} 
\quad . 
\end{eqnarray} 
$D=-i{\vec \partial}_{x_{1}}/2M$ is the Schr\"odinger operator or in a relativistic 
framework the Dirac operator ($D=i\gamma_\mu\partial_{x_{1}}-M$) 
and $\Sigma^{<,>,\pm}$ are the  
self energies. The introduction of retarded and advanced  
functions allows to interpret the real part of the retarded self energy as 
a mean field while the imaginary part describes the  
absorption or finite life times of quasi particles (dressed nucleons)  
\cite{btm90}. The self-energy $\Sigma$ contains all higher order correlations 
and couples the one-body kinetic equation (\ref{TP1}) to the corresponding 
equations for the two- and 3-body densities and so forth. This requires to 
truncate the Dyson-Schwinger hierarchy which is usually done at the two-body 
level and leads to the ladder approximation for T-matrix, i.e. the 
Bethe-Salpeter equation. 

The formal structure of the kinetic equation (\ref{TP1}) is 
complex and  one should solve (\ref{TP1}) together with the corresponding  
kinetic equations for $G^{\pm}$ which describe the {\it spectral properties}  
of the phase space distribution. Simultaneously  
the self-energies should be derived for arbritrary non-equilibrium situations \cite{btm90}. 
A solution of the full self-consistency problem has not yet achieved. 
In praxis one applies further approximations.  
The most important ones are the {\it gradient expansion} (a semi-classical  
approximation to first order in $\hbar$) and the 
{\it quasi-particle approximation} which sets the  
particles on mass shell. The result is a Boltzmann-type transport equation,  
which is known as the Boltzmann-Uheling-Uhlenbeck (BUU) transport equation \cite{bg88}. 
In its relativistic form the (R)BUU equation reads 
\begin{eqnarray} 
&&  \left[ \left( m^{*}_D \p_{x}^\mu m^{*}_D - k^{*\nu} \p_{x}^\mu k^{*}_{\nu} \right) \p^{k}_\mu  
- \left( m^{*}_D \p_{k}^\mu m^{*}_D - k^{*\nu} \p_{k}^\mu k^{*}_{\nu} \right) \p^{x}_\mu  
\right] f 
\nonumber\\
&&=  \frac{1}{2(2\pi)^8} \int \frac{d^3 k_{2}}{E^{*}_{k_{2}}} \frac{d^3 k_{3}}{E^{*}_{k_{3}}} 
             \frac{d^3 k_{4}}{E^{*}_{k_{4}}}~ W~
\delta^4 \left(k + k_{2} -k_{3} - k_{4} \right)    
\nonumber\\
&&\times  \Big[ \: f_3 f_4 \ls 1-  f \rs \ls 1- f_2 \rs -  
  f f_2 \ls 1- f_3 \rs \ls 1- f_4 \rs \: \Big]  
\label{TP5}  
\quad , 
\end{eqnarray} 
which describes the phase space evolution of the $1$-particle distribution  
$f({\bf x},{\bf k},t)$ under the influence of the mean field (which enters  
via the real part of the self energy, i.e. via $m^{*}_D=M-\Sigma_{S}$ and  
$k^{*\mu}=k^{\mu}-\Sigma^{\mu}$) and binary collisions determined by the  
transition amplitude $W=m^{*4}_D |T(kk_2|k_3 k_4)|^2$. Final state Pauli 
blocking is accounted for by the blocking factors $(1-f_i)$ in (\ref{TP5}) 
with $f_i =f({\bf x},{\bf k}_i,t)$.   
The physical parameters entering into the kinetic equation are  
the mean field, i.e. the nuclear EOS, and elementary cross sections for 
$2$-particle scattering processes. Thus one can test the high density
behavior of the nuclear EOS in heavy ion collisions and the in-medium  
modifications of cross sections, which also influence the stopping properties  
of the colliding system. Above the pion threshold where inelastic processes 
start to play an important role the Eq. (\ref{TP5}) becomes a {\it coupled 
channel} problem for nucleonic, nucleon resonance and mesonic degrees of freedom. 
The collision integral, i.e. the right hand side of Eq. (\ref{TP5}) has to 
be extended for the corresponding inelastic and absorptive processes and 
the new degrees of freedom must be propagated in their mean fields. In 
practice the transport equation is solved within the {\it testparticle 
method} which describes the phase space distribution $f$ as an incoherent 
sum of point-like quasi particles \cite{bg88} or static gaussians \cite{gre87}
which propagate on classical trajectories. 
Relativistic formulations of the two methods were developed in \cite{bkm93} 
and \cite{fu95}. 
\subsection{Quantum Molecular Dynamics (QMD)} 
An alternative approach to the kinetic BUU equation is 
Quantum Molecular Dynamics (QMD) \cite{qmd,qmd2,qmd3}.  
QMD is a N-body approach which  
simulates heavy ion reactions on an event by event  
basis taking fluctuations and correlations into account.  
The QMD equations are formally derived  from the assumption that  
the N-body wave function $\Phi$ can be represented as the direct product of  
single coherent states $\Phi = \prod_{i} \phi_{i}$ which are described 
by Gaussian wave packets. Anti-symmetrization is {\it not} taken into 
account. A Wigner transformation yields the corresponding 
phase space representation of $\Phi$. The equations of motion of the many-body  
system are obtained by the variational principle starting  
from the action $S = \int {\cal L}[\Phi, \Phi^{*}]$  
(with the Lagrangian functional  
${\cal L} = <\Phi|i\hbar \frac{d}{dt}-H|\Phi>$).  
The Hamiltonian $H$ contains a kinetic contribution and mutual two-body interactions  
$V_{ij}$. The variational principle leads finally to classical equations of motion  
for the generalized coordinates ${\bf q}_{i}$ and ${\bf k}_{i}$ of the
Gaussian wave packets
\begin{eqnarray} 
\dot{{\bf q}}_{i}  &=&  \frac{{\bf k}_{i}}{m}  
+ \nabla_{{\bf k}_{i}} \sum_{j\neq i}<V_{ij}> = \nabla_{{\bf k}_{i}}<H> 
~~, \nonumber \\
\dot{{\bf k}}_{i} &=&   
- \nabla_{{\bf q}_{i}} \sum_{j\neq i}<V_{ij}> = \nabla_{{\bf q}_{i}}<H> 
\nonumber 
\quad . 
\end{eqnarray} 
The two-body interaction $V_{ij}$ can be e.g. taken from BHF 
calculations \cite{qmd3} or from local Skyrme forces which are 
usually supplemented by an empirical momentum dependence in order 
to account for the energy dependence of the optical nucleon-nucleus 
potential \cite{qmd}. Binary collisions are 
treated in the same way as in BUU models. 
Furthermore there exist relativistic extensions, i.e. RQMD 
and the UrQMD model which has been developed to simulate heavy ion  
collisions at ultra-relativistic energies \cite{qmd2,rqmd}. 
\subsection{Antisymmetrized Molecular Dynamics (AMD/FMD)} 
An extension of QMD, in particular designed for low energies, 
are the Antisymmetrized Molecular Dynamics (AMD) \cite{ono92} 
and Fermionic Molecular Dynamics  (FMD) approaches \cite{feldmeier}. 
In contrast to conventional QMD, 
the interacting system is represented by an {\it antisymmetrized}  
many-body wave function consisting of single-particle states which  
are localized in phase space. The equations of motion for the parameters  
 characterizing the many-body state (e.g. position, momentum, width and spin  
 of the particles) are derived from a quantum variational principle.  
 The models are designed to describe ground state properties of nuclei  
 as well as heavy ion reactions at low energies (see also Chap. I.2).
\subsection{Off-shell transport} 
Essential for the validity of the classical equations of motion is the 
quasi-particle approximation (QPA) which assumes that the spectral
strength of a hadron 
is concentrated around its quasi-particle pole. 
Particle widths can, however, dramatically change in a
dense hadronic environment. To first order in density the in-medium width 
of a hadron in nuclear matter can be estimated by the collision width 
$\Gamma^{\rm tot}=\Gamma^{\rm vac} + \Gamma^{\rm coll}, 
\Gamma^{\rm coll} = \gamma v \sigma \rho_B$ 
with $v$ the hadron velocity relative to the surrounding matter and $\sigma$
the total hadron-nucleon cross section. A consistent 
treatment of the off-shell dynamics, i.e. a solution of 
the quantum evolution equations for the correlation 
functions $G^{<,>}$ has up to now only been performed for 
toy models and simplified geometries \cite{dani84,koehler95} or 
in first order gradient approximation leading to an extended 
quasi-particle picture \cite{MLS00}. Comparing the nonlocal 
extension of BUU
with standard simulations a visible effect of nonlocal correlations 
is seen and a better agreement with measured  charge 
density distributiond \cite{MLNCCT01} or particle spectra \cite{MSLKKN99} 
due to the virial corrections has been found. To develop 
a consistent lattice quantum transport for non-uniform systems and realistic 
interactions will be one of the future challenges in theoretical 
heavy ion physics. 

On the other hand, substantial progress has been  made in the recent 
years to map 
part of the off-shell dynamics on a modified test-particle formalism 
\cite{cassing00,lehr00}. This allows to apply 
off-shell dynamics, although in a simplified form, to the complex 
space time evolution of a heavy ion reaction. The present knowledge 
of off-shell matrix elements is, however, rather limited and theoretical  
investigations are scarce \cite{offcs}. The off-shell T-matrix 
has been used in order to calculate the duration and non-locality 
of a nucleon-nucleon collision \cite{MLSK98}. 
The question to what degree a depletion of the Fermi surface due to 
particle-hole excitations and the high momentum tails of the nuclear 
spectral functions will affect subthreshold particle 
production is not so obvious to answer. The high momentum tails 
correspond to deeply bound states which are off-shell and to treat 
such states in a standard transport 
approach like on-shell quasi-particles 
would violate energy-momentum conservation. Energy-momentum 
conservation can be achieved consistently by the nonlocal 
kinetic theory \cite{LSM97} taking into account first order 
off-shell effects. The contribution 
of the nuclear short-range correlations to subthreshold $K^+$ production 
in p+A reactions have e.g. been estimated in \cite{saturne96}. 
The removal energy 
for a high momentum state compensates the naively expected energy 
gain and the short-range 
correlations do therefore not significantly contribute to subthreshold 
particle production \cite{saturne96}. 
The situation changes, however, when the medium 
is heated up and high momentum particles become on-shell or when 
the spectral distributions of the produced hadrons themselves 
are broadened.   
\section{Constraints from heavy ion collisions} 
\subsection{Flow and stopping} 
On of the most important observable to constrain the nuclear forces  
and the underlying EOS at supra-normal densities  
is the collective nucleon flow \cite{norbert99}.  
It can be characterized in terms of anisotropies of the azimuthal  
emission pattern. Expressed 
in terms of a Fourier series  
\begin{equation} 
\frac{dN}{d\phi} \propto 1 +  2v_1cos(\phi)  
+2 v_2cos(2\phi) + \dots  
\end{equation} 
this allows a transparent interpretation of the coefficients  
$v_1$ and $v_2$. The dipole term $v_1$  
arises from a collective sideward deflection of  
the particles in the reaction plane and characterizes the  
transverse flow in the reaction plane. The second harmonics  
describes the emission pattern perpendicular to the reaction  
plane. For negative $v_2$ one has a preferential out-of-plane 
emission.  The phenomenon of an out-of-plane 
enhancement of particle emission at midrapidity is  
called {\it squeeze-out}.  
 
The transverse flow $v_1$ has been found to be sensitive to the EOS 
and, in particular in peripheral reactions, to the momentum dependence  
of the mean field \cite{dani00,gaitanos01}. The elliptic flow $v_2$,  
in contrast, is very sensitive to the maximal compression reached in 
the early phase of a heavy ion reaction. The cross over from 
preferential in-plane flow $v_2 < 0$ to preferential  
out-off-plane flow  $v_2 > 0$ around 4-6 AGeV has also led to 
speculations about a phase transition in this energy region which  
goes along with a softening of the EOS \cite{eflow}.   
\begin{figure}[h] 
\resizebox{0.98\columnwidth}{!}{\includegraphics{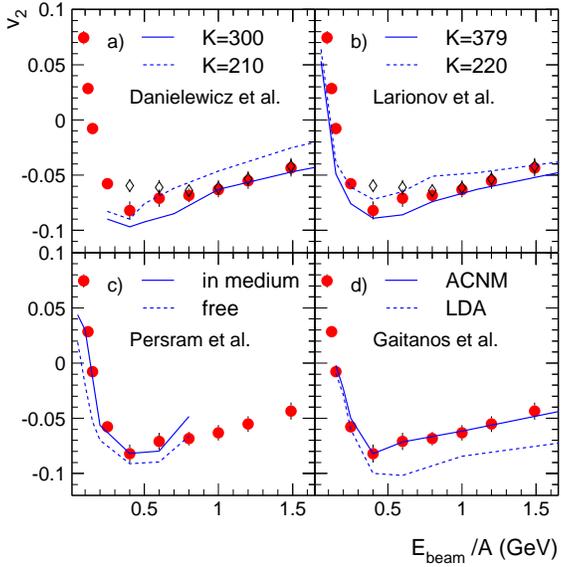}}
\caption{Elliptic flow excitation function at SIS energies.  
Various theoretical studies using different EOS's ((a),(b)),  
or different cross sections (c) or DBHF mean fields in the  
LDA approach and including further non-equilibrium effects (ACNM) 
are compared to FOPI data (symbols). Figure  
is taken from \protect\cite{fopi05}. 
} 
\label{v2_fig} 
\end{figure} 
The present situation between theory and experiment is illustrated in  
Fig. \ref{v2_fig} (from \cite{fopi05}). The BUU studies from 
Danielewicz et al. and the Giessen group (Larionov et al.) investigated  
the EOS dependence while Persram et al. find a sensitivity of  $v_{2}$ 
on the medium dependence of the NN cross sections. Finally,  
non-equilibrium effects have been investigated at the level of the effective 
interaction in \cite{gaitanos01,fuchs03}. It has been found that the 
 local phase space anisotropies of the pre-equilibrium stages of the 
reactions reduce the repulsion of the mean field and soften the 
corresponding EOS which allows a good description of the $v_2$ data 
using microscopic DBHF mean fields 
(Gaitanos et al.). However, Fig. \ref{v2_fig} also demonstrates 
that  $v_{2}$ is generated by the interplay of the mean field and binary 
collisions which makes it difficult to extract exclusive information 
on the EOS from the data. Here certainly further going studies are 
required. 

The next figure \ref{eos_pressure_fig}, based on the studies of 
Danielewicz \cite{dani02}, summarizes the status obtained 
within this model in terms of a  
band that represents the constraints from  
collective flow data. It based on a compilation from 
the analysis of sideward and  elliptic anisotropies, 
ranging from low SIS ($E_{\rm lab} \simeq 0.2\div 2$ AGeV)  
up to top AGS energies ($E_{\rm lab} \simeq 2\div 11$ AGeV),  
has been studied in \cite{dani02}. The conclusion from  
this study was that, both, super-soft equations of state (K=167 MeV)  
as well as hard EOSs (K$>$300 MeV) are ruled out by  data. 
At SIS energies  existing flow data are consistent with a soft  
EOS \cite{hombach99,dani00}, e.g. the The soft Skyrme EOS.  
In the models used by Danielewicz et al.  
\cite{dani00,dani02} sideward flow favors a rather soft  
EOS with K=210 MeV  
while the development of the elliptic flow requires slightly  
higher pressures. The BHF and variational calculations  
including 3-body-forces\protect\footnote{For the  
BHF + 3-BF calculation the pressure shown in Fig.  
\ref{eos_pressure_fig} has been determined from  
the parameterization given in \protect\cite{zuo04} which is  
based on the Urbana IX 3-BF different to that used in \protect\cite{zuo02}.}  
fit well into the constrained area up to 4$\rho_0$. At higher  
densities the microscopic EOSs, also DBHF, tend to be too  
repulsive. 
\begin{figure}[h] 
\resizebox{0.92\columnwidth}{!}{\includegraphics{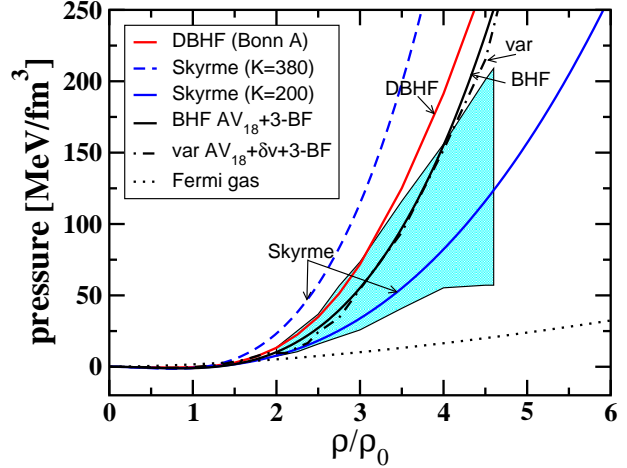}}
\caption{Constraints on the nuclear EOS from heavy ion flow data.  
The shaded area shows the pressure-density which is compatible with  
heavy ion flow data according the analysis on \protect\cite{dani02}.  
The equations-of-state from the models  
shown in Fig. \protect\ref{dbeos_fig} are displayed. 
} 
\label{eos_pressure_fig} 
\end{figure} 
However, conclusions from flow data are  
generally complicated by the interplay of the compressional part  
of the nuclear EOS and the momentum dependence of the nuclear  
forces. A detailed comparison to  $v_1$ and $v_2$ data from  
FOPI \cite{andronic99,stoicea04} and KaoS \cite{brill96}  
for $v_1$ and $v_2$ below 1 AGeV favors  
again a relatively soft EOS with a momentum dependence 
 close to that obtained from microscopic DBHF calculations  
\cite{dani00,gaitanos01,lca}. In Fig.  
\ref{eos_pressure_fig} the microscopic DBHF EOS (K=230 MeV)  
lies at the upper edge of the boundary, but is still consistent   
in the density range tested at SIS energies,  
i.e. up to maximally 3 $\rho_0$. This fact is further  
consistent with the findings of Gaitanos et al.  
\cite{gaitanos01,lca} where a good  
description of $v_1$ and $v_2$ data at energies between  
0.2 and 0.8 AGeV has been found in RBUU calculations based  
on DBHF mean fields. As pointed out in 
\cite{gaitanos01,fuchs03,lca} it is thereby essential  
to account for non-equilibrium effects and the momentum  
dependence of the forces which softens the EOS compared to  
the equilibrium case (shown in  Fig. \ref{eos_pressure_fig}). 

As can be seen from Fig. \ref{v2_fig}, not only the nuclear EOS,  
but also the cross sections for elementary $2$-particle scattering  
influences the collective dynamics, in particular the degree of  
stopping and hence the maximal compression achieved in the fireball  
region. A challenge is in this context the quantitative understanding of 
the recently observed strong correlations between maximal side flow $v_1$ 
and maximal stopping in both excitation functions \cite{reisdorf04} (see 
Chap. I.5 by W. Reisdorf et al.).  
Most collective flow analyses performed so far have were based on free 
cross sections, which works  astonishingly well from a 
practical point of view. However, within a consistent picture one should  
treat the in-medium effects in both, real (nuclear EOS) and   
the imaginary part (cross sections) of the interaction on the same 
footing. Many-body calculations (BHF/DBHF)  
predict an essential reduction of the elastic NN cross section with increasing baryon  
density \cite{qmd3,lima93,offcs} and in \cite{larionov01} a similar 
reduction was proposed for the inelastic channels in oder to 
describe pion multiplicities in the 1-2 AGeV region. 
One can therefore expect observable signals in heavy ion collisions.  
In fact, recent QMD studies of stopping and transparency observables  
have shown that the data can be reproduced when the free cross section  
is reduced by a factor of 0.5 \cite{hong05}. These findings are supported 
by transport calculations using microscopic in-medium cross sections 
\cite{gait05,danic}.  Therefore, for a reliable extraction  
of the high density nuclear EOS one should account for in-medium effects not  
only in the potential, but also in the cross sections. 
\subsubsection{Isospin dependence of the EOS} 
Another important aspect of heavy ion collisions is the investigation  
of the density dependence of the EOS for asymmetric matter. There exist  
abundent studies on this sector, either non-relativistically or  
relativistically.  

The momentum dependence of the isovector potential, Fig. \ref{uopt_fig}, 
which is also closely related to the  proton/neutron mass splitting 
of both, the non-relativistic $m^*_{NR}$ and the Dirac $m^*_{D}$ effective 
mass, is one of the key questions which can be addressed by  
nuclear reactions induced by neutron-rich nuclei at RIA energies. 
Transverse and elliptic flow pattern as well 
$p/n$ rapidity dsitributions have been suggested 
as possible observables to investigate the 
momentum dependence and the $p/n$ mass splitting \cite{bao2,bao04,rizzo}.

Promising observables to pin down the density dependence of the 
symmetry energy are the iso-scaling 
behavior of fragment yields and the isospin diffusion in 
asymmetric colliding systems. In both cases 
recent NSCL-MSU data in combination with 
transport calculations are consistent with a value of 
$E_{\rm sym} \approx 31 $ at $\rho_0$ and 
rule out extremely " stiff " and " soft " density 
dependences of the symmetry energy \cite{chen05,yennello05} (see also Chap. III.2). 
The same value has been extracted \cite{khoa05} 
from low energy elastic and (p,n) charge exchange 
reactions on isobaric analog states, i.e. p($^{6}He,^{6}Li^*$)n measured at the HMI. 
Such a behavior is also consistent with the predictions from many-body theory 
\cite{akmal98,dalen04}. Also the $p/n$ ratio at mid-rapidity has been found to 
be sensitive to the high density behavior of nuclear symmetry energy \cite{bao05}.

In relativisitic approaches 
large attractive scalar and repulsive vector fields 
are required by Dirac phenomenology in order to describe simultaneously 
the central potential and the strong spin-orbit force in finite nuclei 
\cite{rmf,eft1,finelli}. The situation is, however, 
less clear in the iso-vector sector. There exists  
different possibilities to reproduce the same value of the  
$a_{4}$ coefficient (\ref{esym}): (a) by only an iso-vector 
vector $\rho$ field like in most RMF models (NL3 etc.), 
or (b) by accounting for an additional iso-vector 
scalar $\delta$ field.  
Due to competing effects between attractive (scalar $\delta$) and  
repulsive (vector $\rho$) both alternatives can be fitted to the same empirical  
$a_{4}$ parameter, but the inclusion of $\delta$ field leads to an 
essentially different high density behavior of the 
symmetry energy \cite{liu}. The scalar $\delta$ field is suppressed at high densities,  
whereas the vector field is proportional to the baryon  
density which makes the symmetry  stiffer energy at supra-normal densities.  
Recent transport studies have shown that these subtle relativistic effects  
can be observed in the intermediate energy range by means of  
collective isospin flow, particle ratios and imbalance ratios of different  
particle species (protons, neutrons, pions and kaons)
\cite{liu,rizzo,baran04}. 
However, due  to the lack of precise experimental data no 
definitive conclusions could be made so far. 
\subsection{Particle production} 
\subsubsection{Pions} 
With the start of the first relativistic heavy ion programs the  
hope was that particle production would provide a direct experimental  
access to nuclear EOS \cite{stock86}. At two times  
saturation density  which is reached in the participant zone of  
the reactions without additional compression the difference between  
the  soft and hard EOS shown in Fig. \ref{dbeos_fig} is about  
13 MeV in binding energy. If the matter is compressed up to 3$\rho_0$  
the difference is already $\sim 55$ MeV. It was expected that the   
compressional energy should be released into the creation of  
new particles, primarily pions, when the matter expands \cite{stock86}.  
However,  
pions have large absorption cross sections and they turned out not to  
be suitable messengers of the compression phase. They undergo several  
absorption cycles through nucleon  resonances ($N\pi \leftrightarrow \Delta$)  
and freeze out at final stages of the reaction  
and at low densities. Hence pions loose most of  
their knowledge on the compression phase and are not  
really sensitive probes for stiffness of the EOS \cite{senger99}.  
However, they carry information on the isotopic composition of the matter  
which is to some extent conserved until freeze-out.  
The final $\pi^-/\pi^+$ ratio was found to be sensitive  
to the initial $n/p$ composition of the matter which, on the other hand,  
is influenced by the isospin dependence of the nuclear forces  
\cite{uma97,bao05b}. In \cite{gaitanos04b} a reduction of  
the $\pi^{-}/\pi^{+}$-ratio was found when the $\delta$-meson  
was included in the RMF approach. The effects are, however, moderate, i.e.  
at the 10-20\% level, and most pronounced at extreme phase space  
regions, e.g. at the high energy tails of $p_t$ spectra 
\cite{bao05b,gaitanos04b,urqmd05}. Systematic  
measurements, e.g. from the FOPI Collaboration may help to  
constrain the isospin dependence by pionic observables.  
\subsubsection{Kaons} 
\begin{figure}[h] 
\resizebox{0.98\columnwidth}{!}{\includegraphics{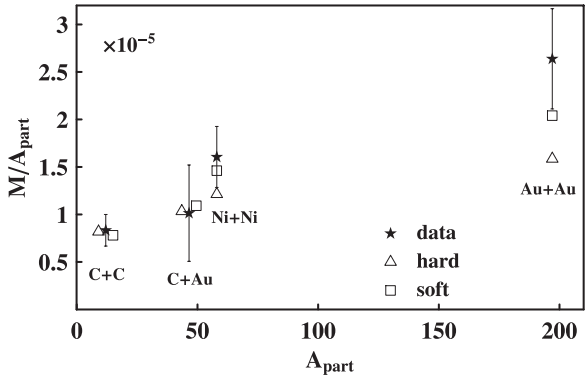}}
\caption{$K^+$ multiplicities in inclusive C+C,~Ni+Ni,~Au+Au and  
C+Au reactions at 1 AGeV. QMD calculations using a hard/soft nuclear EOS are  
compared to KaoS data \protect\cite{kaos05}. Figure is taken  
from   \protect\cite{kaos05}. 
} 
\label{fig_aparteos} 
\end{figure} 
After pions turned out to fail as suitable messengers,  
$K^+$ mesons were suggested as promising tools to probe the nuclear  
EOS, almost 20 years ago \cite{AiKo85}.  
The cheapest way to produce a $K^+$  meson  
is the reactions $NN\longrightarrow N\Lambda K^+$  which has a  
threshold of $E_{\rm lab} =1.58$ GeV kinetic energy for the incident  
nucleon. When the  
incident energy per nucleon in a heavy ion reactions is below this  
values one speaks about {\it subthreshold} kaon production. Subthreshold  
kaon production is in particular interesting since it ensures that the  
kaons originate from  the high density phase of the reaction. The missing  
energy has to be provided either by the Fermi motion of the nucleons or  
by energy accumulating multi-step reactions. Both processes exclude significant 
distortions from surface effects if one goes sufficiently far below  
threshold. In combination with the long mean free path subthreshold $K^+$  
production is an ideal tool to probe compressed nuclear matter in relativistic  
heavy ion reactions. 
 
Already in the first theoretical  
investigations by transport models it was noticed the $K^+$  
yield reacts rather sensitive on the EOS  
\cite{huang93,hartnack94,li95b}.  
Both, in non-relativistic QMD calculations  
based on soft/hard Skyrme forces \cite{huang93,hartnack94,fuchs97b}  
and in RBUU \cite{li95b,giessen94} with soft/hard versions  
of the (non-linear)  
$\sigma\omega$--model the $K^+$ yield was found to be  
about a factor 2--3 larger when a soft EOS  
is applied compared to a hard EOS. At that time the available  
data favored a soft equation of state \cite{hartnack94,li95b,giessen94}.  
However, at that stage the  
theoretical calculations were still burdened with large uncertainties.  
First of all, it was noticed \cite{huang93,hartnack94} that  
the influence of the repulsive momentum dependent  
part of the nuclear interaction leads to  
a strong suppression of the kaon abundances which made a quantitative  
description of the available data more difficult. Moreover, at that  
time the pion induced reaction channels $\pi B\longrightarrow YK^+ $  
have not yet been taken into account. These additional channels  
which contribute up to $30\div 50\%$ to the  
total yield enabled to explain the measured yields with  
realistic momentum dependent interactions \cite{fuchs97b,brat97}.  
A breakthrough was achieved when the  COSY-11 Collaboration  
measured the $pp \longrightarrow p K^+ \Lambda$ reactions at  
threshold \cite{cosy11} which constrains the strangeness production cross  
sections $NN \longrightarrow N K^+ Y$  
Within the last decade the KaoS Collaboration has performed systematic  
measurements of the $ K^+$ production far below threshold  
\cite{senger99,kaos94,sturm01}. Based on the new data situation,  
the question  
if valuable information on the nuclear EOS can be extracted  
has been revisited and it has been  
shown that subthreshold  $K^+$ production provides indeed a suitable and  
reliable tool for this purpose \cite{fuchs05,fuchs01,hartnack01}.

Fig.  \ref{fig_aparteos} compares measured the $K^+$ multiplicities as a  
function of participating nucleons  
$A_{\rm part}$ in Au+Au, Ni+Ni, C+Au and C+C reactions at 1 AGeV to  
QMD calculation using a soft/hard momentum dependent Skyrme force  
\cite{kaos05}. This figure demonstrates thereby the interplay  
between $A_{\rm part}$, system size and EOS.  
A significant dependence of the kaon multiplicities on  
the nuclear EOS requires a large amount of collectivity which is easiest  
reached in central reactions of heavy mass systems. Consequently, the  
EOS dependence is most pronounced in central Au+Au reactions. Also  
in Ni+Ni effects are still sizable while the small C+C  
system is completely insensitive on the nuclear EOS even in most central  
reactions. The data available for Au+Au and Ni+Ni  support the soft EOS.  
Interesting is in this context the asymmetric C+Au system: Though in  
central C+Au reactions the number of participants is comparable to  
Ni+Ni the $K^+$ yield does not depend on the EOS. This indicates again  
that a sensitivity on the  EOS is not only a question of  $A_{\rm part}$ but  
of the compression which can be reached by the colliding system.  
\begin{figure}[h] 
\resizebox{0.98\columnwidth}{!}{\includegraphics{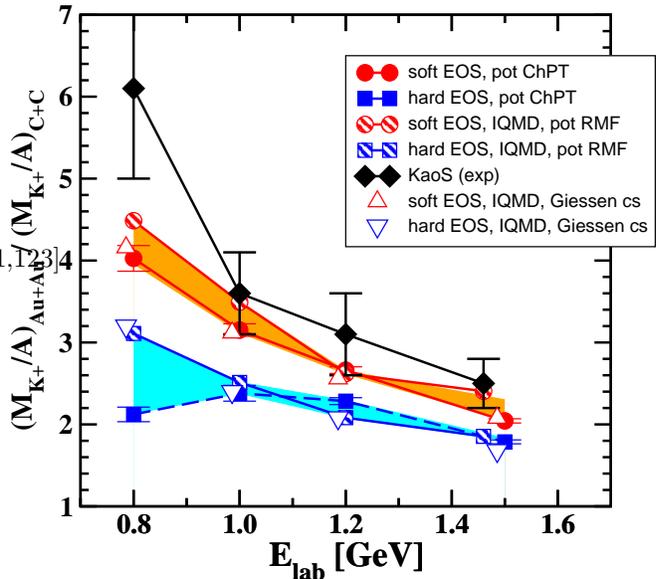}}
\caption{Excitation function of the ratio $R$ of $K^+$  
multiplicities obtained in inclusive Au+Au over C+C  
reactions. QMD  \protect\cite{fuchs01} and IQMD  
calculations \protect\cite{hartnack01} are compared to  
the KaoS data \protect\cite{sturm01}. The shaded area indicates 
thereby the range of uncertainty in the theoretical models.  
In addition IQMD results based on an alternative set of  
elementary $K^+$ production cross sections are shown. 
}
\label{fig_ratio_1} 
\end{figure} 
The next step is to consider now the energy dependence of the EOS  
effect. It is expected to be most pronounced most far below threshold because  
there the highest degree of collectivity, reflected in multi-step collisions, 
is necessary to overcome the production thresholds.   
The effects become even more evident when the ratio $R$ of the  
kaon multiplicities obtained in Au+Au over C+C  
reactions (normalized to the corresponding mass numbers) is built  
\cite{fuchs01,sturm01}. Such a ratio has moreover the advantage that  
possible uncertainties which might still exist in the  
theoretical calculations should cancel out  
to large extent. This ratio is shown in Fig. \ref{fig_ratio_1}.  
Both, soft and hard EOS, show an increase of $R$ with decreasing energy. 
However, this increase is much less  
pronounced when the stiff EOS is employed.  
The strong increase of $R$ can be directly related to  
higher compressible nuclear matter. The comparison to the experimental  
data from KaoS \cite{sturm01} where the increase of $R$ is even more  
pronounced strongly favors a soft equation of state.  
These findings were confirmed by independent IQMD transport calculations of  
the Nantes group \cite{hartnack01}. 
Both, QMD and IQMD included also a repulsive kaon-nucleon potential as  
predicted by chiral perturbation theory \cite{fuchs05}. The  
shaded area in the figure can be taken as the existing range of  
uncertainty in the theoretical model description of the considered  
observable. To estimate the stability of the conclusions   
the IQMD calculations have been  repeated with an alternative  
set of $N\Delta; \Delta\Delta \mapsto NYK^+$ cross sections 
\footnote{Cross section which involve $\Delta$ resonances in the initial  
of final states are not constrained by measurements.}  
which are almost one order of magnitude smaller than those used  
originally, but the EOS dependence remained stable \cite{hartnack05}.   
\section{Constraints from neutron stars}
Measurements of ``extreme'' values,
like large masses or radii, huge luminosities etc.\
as provided by compact stars offer good opportunities to gain deeper insight
into the physics
of matter under extreme conditions. 
There has been substantial progress in recent time from the astrophysical 
side. 

The most spectacular observation was probably the 
recent measurements on PSR J0751+1807, a millisecond pulsar in a binary 
system with a helium white dwarf secondary, which implies
a pulsar mass of $2.1\pm0.2\left(^{+0.4}_{-0.5}\right) {\rm M_\odot}$
with $1\sigma$ ($2\sigma$) confidence~\cite{NiSp05}.
Therefore a reliable EOS has to describe neutron star
(NS) masses of at least $1.9~ {\rm M}_\odot$ ($1\sigma$) in a strong,
or  $1.6~ {\rm M}_\odot$ ($2\sigma$) in a weak interpretation.
This  condition limits the softness of EOS in NS matter. One might 
therefore be worried about an apparent contradiction between the 
constraints derived from neutron stars and those from heavy ion 
reactions. While heavy ion reactions favor a soft EOS, PSR J0751+1807  
requires a stiff EOS. The 
corresponding constraints are, however, complementary rather than contradictory. 
Intermediate energy heavy ion reactions, e.g. subthreshold kaon production, 
constrains the EOS at densities up to $2\div3~\rho_0$ while the maximum NS 
mass is more sensitive to the high density behaviour of the EOS. Combining the 
two constraints implies that the EOS should be {\it soft at moderate 
densities and  stiff at high densities}. Such a behaviour is predicted 
by microscopic many-body calculations (see Fig. \ref{dbeos_fig}). DBHF, BHF 
or variational calculations lead typically the maximal NS masses between  
$2.1\div 2.3~M_\odot$ and are therefore in accordance with PSR J0751+1807 \cite{klaehn06}. 

There exist several other constraints on the nuclear EOS which can 
be derived from observations of  compact stars, see e.g. \cite{klaehn06,steiner05,steiner05b}. 
Among these the one of the 
most promising constraints is the Direct Urca (DU) process which 
is essentially driven by the proton fraction inside the NS \cite{lattimer91}. 
DU processes, e.g. the neutron $\beta$-decay $n\to p+e^-+\bar\nu_e$,
are very efficient regarding their neutrino production,
even in superfluid NM ~\cite{BlGrVo04,KoVo05},
and cool NSs too fast to be in accordance
with data from thermal observable NSs.
Therefore one can suppose that
no DU processes should occur
below the upper mass limit
for ``typical'' NSs, i.e.
$M_{DU}\ge 1.5~M_\odot$ ($1.35~M_\odot$ in a weak interpretation).
These limits come from a population synthesis of young,
nearby NSs~\cite{Popov:2004ey} and masses of NS binaries  ~\cite{NiSp05}.

\section{Summary and outlook}
The quest for the nuclear equation of state is one of the 
longstanding problems in physics which has a more than 50 years 
history in nuclear structure. Since about 30 years one tries 
to attack this question with heavy ion reactions. The exploration 
of the limits of stability, i.e. the regimes of extreme isospin 
asymmetry, is  a relatively new field with rapidly growing 
importance in view of the forthcoming generation of radioactive 
beam facilities. 

The status of theoretical models which make predictions for the EOS 
can roughly be summarized as follows: phenomenological density 
functionals such as Skyrme, Gogny or relativistic mean field models 
provide high precision fits to the nuclear chart but extrapolations 
to supra-normal densities or the limits of stability are highly 
uncertain. A more controlled way provide effective field theory 
approaches which became quite popular in recent time. Effective 
chiral field theory allows e.g. a systematic generation of two- 
and many-body nuclear forces. However, these approaches are low 
momentum expansions and when applied to the nuclear many-body 
problem, low density expansions. Ab initio calculations for the 
  nuclear many-body problem such as variational or Brueckner 
calculations have reached a high degree of sophistication and 
can serve as guidelines for the extrapolation to the regimes of 
high density and/or large isospin asymmetry. Possible future 
devellopments are to base such calculations on modern EFT 
potentials and to achieve a more consistent treatment of 
two- and three-body forces. 

If one intends to constrain these models by nuclear reactions 
one has to account for the reaction dynamics by semi-classical 
transport models of a Boltzmann or molecular dynamics type. 
Suitable observables which have been found to be sensitive on 
the nuclear EOS are directed and elliptic 
collective flow pattern and particle production, 
in particular kaon production, at higher energies. Heavy ion 
data suggest that the EOS of symmetric nuclear matter shows a soft 
behavior in the density regime between one to about three times 
nuclear saturation density, which is consistent with the 
predictions from many-body calculations.  Conclusions 
on the EOS are, however, complicated by the interplay between the 
density and the momentum dependence of the nuclear mean field. 
Data which constrain the isospin dependence of the mean field are 
still scare. Promising observables are isospin diffusion, iso-scaling 
of intermediate mass fragments and particle ratios 
($\pi^+/\pi^-$ and eventually $K^+/K^0 $). Here the situation 
will certainly improve when the forthcoming radioactive beam 
facilities will be operating. This will also allow to measure 
the optical isospin potential in p+A and A+A reactions and to obtain more 
information of the symmetry energy and the proton/neutron 
mass splitting in asymmetric matter. From the theoretical side 
it will be unavoidable to invest significant efforts towards 
the devellopment of quantum transport models with consistent off-shell 
dynamics.

 
We would like to thank K. Morawetz, T. Gaitanos and M. di Toro 
for fruitful discussions.

\end{document}